\documentclass[aps,amssymb,amsmath,preprint,showpacs]{revtex4}
\usepackage{graphicx}
\newcommand{\ket}[1]{\left|{#1}\right\rangle}
\newcommand{\bra}[1]{\left\langle{#1}\right|}

\newcommand{\expect}[3]{\left\langle{#1}\right|{#2}\left|{#3}\right\rangle}
\newcommand{\aver}[1]{\left\langle{#1}\right\rangle}
\newcommand{\added}[2]{\left|{#1},{#2}\right\rangle}
\newcommand{\inner}[2]{\left\langle{#1}|{#2}\right\rangle}
\newcommand{\beq}{\begin{equation}}
\newcommand{\eeq}{\end{equation}}
\begin{document}
\title{Wave packet dynamics of entangled two-mode states}

\author{C. Sudheesh, S. Lakshmibala, and
V. Balakrishnan}
\email{sudheesh,slbala,vbalki@physics.iitm.ac.in}
\address{
 Department of Physics, Indian Institute of Technology
  Madras, Chennai 600 036, India}

\begin{abstract}
We consider a model Hamiltonian describing the interaction of a
single-mode radiation field with the atoms of a nonlinear medium, and 
study the dynamics of entanglement for specific non-entangled 
initial states of interest: namely, those in which the field mode is
initially in a Fock state, a coherent state, or a photon-added
coherent state. The counterparts of near-revivals and fractional
revivals are shown to be clearly identifiable in the entropy of
entanglement. The ``overlap fidelity'' of the system is another such
indicator, and its behaviour corroborates that of the entropy of
entanglement in the vicinity of near-revivals. The expectation values
and higher moments of suitable quadrature variables are also examined,
with reference to possible squeezing and higher-order squeezing.    

\end{abstract}

\vspace{2pc}

\maketitle

\section{Introduction}
\label{introdn}

A problem of considerable interest in 
quantum dynamics is that of the identification of  
signatures of non-classical effects in the temporal behaviour of   
quantum mechanical expectation values 
in nonlinear systems. The dynamics of a quantum wave packet 
governed by a nonlinear 
Hamiltonian provides adequate scope for such an investigation to be 
carried out, as a wide variety of non-classical effects such as 
revivals and fractional revivals \cite{robi}, as well as squeezing, are 
displayed by the wave packet as it evolves in time.

While a generic 
initial wave packet $\ket{\psi(0)}$ governed by a nonlinear 
Hamiltonian spreads rapidly during its evolution,  
it could  return to its original 
state (apart from an overall phase)  at multiples of a 
revival time $T_{\rm rev}$, under 
certain conditions. This is signalled by 
the return of 
the overlap $C(t) = 
|\inner{\psi(0)}{\psi(t)}|^2$  
to its initial 
value of unity at $t = n T_{\rm rev}$.      
Further, at specific instants of time in between 
two successive revivals, fractional
revivals of the wave packet may occur. This is 
characterised by the splitting
up of the initial wave packet into a number of 
spatially distributed sub-packets,
each of which is similar to
the original wave packet. Both revivals and fractional revivals of a 
wave packet arise due to very specific quantum interference properties 
between the basis states 
comprising the original wave packet \cite{tara}.

It is evident that, at exact revivals of any initial state,  quantum
mechanical expectation values of observables return to their
initial values. Distinctive signatures of different     
fractional revivals show up in the time dependence of the 
higher moments of appropriate operators. By tracking the 
time evolution of various 
moments of  certain operators, selective 
identification of 
different fractional revivals can be achieved \cite{sudh1}.
Since the initial state of the system 
turns out to play a crucial role 
in determining its 
subsequent dynamics and the non-classical features it exhibits, these 
signatures also help 
assess the degree of coherence of the initial state 
\cite{sudh2}. Further, it has been shown that 
the squeezing and higher-order squeezing 
properties of certain quadrature variables in the neighbourhood of 
a fractional revival of the wave packet provide 
quantifiable 
measures of the departure from perfect coherence of the initial 
state \cite{sudh3}. The studies in Refs. \cite{sudh1}-\cite{sudh3} 
have been carried out in the context of the propagation of 
a single-mode electromagnetic field 
in a Kerr-like medium, modelled by
the Hamiltonian $H =  \hbar\chi \,a^{\dagger 2}a^2$,  
where $a\,,\,a^\dagger$ are photon annihilation
and creation operators, and $\chi\,\,(> 0)$ 
represents the susceptibility of the nonlinear
medium. (We shall refer to this 
as the case of ``single-mode'' dynamics in what follows.)  
The initial state of the field has been 
taken to be a member of the  
family of photon-added coherent states \cite{agar},  
as the properties of such states include 
a quantifiable degree of departure
from perfect coherence, sub-Poisson statistics 
(a standard deviation that is asymptotically proportional to 
a power of the mean that is less than 
$\frac{1}{2}$), and phase-squeezing. The standard 
oscillator coherent state (CS) is a limiting case of 
a photon-added coherent state (PACS). 

A similar investigation of the dynamics of
two interacting modes is of special interest, in view of an  
additional phenomenon that occurs in this 
case---namely, entanglement. 
Interesting aspects 
of the entanglement that arises when an initial  single-mode 
coherent state passes through a nonlinear medium modelled by the 
above-mentioned Hamiltonian, followed by an interaction with a $50\%$
beam splitter, have been discussed by van Enk \cite{enk}. 
Sanz {\it et al.} \cite{sanz} 
have examined the  
non-classical effects that arise in the 
dynamics of two entangled modes 
governed by a nonlinear 
Hamiltonian, in the framework of an exactly 
solvable case: 
two modes of an electromagnetic field interacting in a 
Kerr-like medium. Taking the initial state 
to be a direct product, either of two Fock 
states or of two coherent states,  
the periodic exact revival of these states has been established,  
and the manner in which 
these properties are mirrored in the collapse and revival 
phenomena displayed by the expectation values of 
appropriate observables has 
been investigated. The collapses are 
marked by expectation values 
remaining constant over a certain interval, 
while the revivals are signalled 
by rapid pulsed variations of the expectation 
values. These features are the close analogues  
of those   
displayed in the single-mode case mentioned earlier. It must be noted,
however, that the specific Hamiltonian considered in these studies is
symmetric in the two modes, with identical nonlinear terms and
a symmetric coupling between the modes. As a consequence, the
Hamiltonian is readily diagonalised, and the resultant dynamics
displays a
considerable degree
of regularity, including the occurrence of revivals.

However, this symmetry is not expected to be 
present in a generic Hamiltonian 
governing the interaction of a single-mode field with a nonlinear
medium. The best one can hope for is the 
occurrence of approximate or 
{\it near}-revivals for most values of the 
parameters in the Hamiltonian.     
In addition to  near-revivals and related phenomena, 
several other 
features of interest are exhibited in general by 
the dynamics 
of two interacting modes. 
Our objective in this paper 
is to demonstrate and investigate these.
We will also examine the link between the  extent to which 
revivals occur and the nature of the initial state. 
A suitable Hamiltonian for our purposes
is the one \cite{puri} that  
describes the interaction of   
a single-mode field  with the atoms of the nonlinear 
medium  through which 
it propagates. The latter is modelled 
by an anharmonic oscillator.
The important point to note is that this Hamiltonian is not 
symmetric in the two interacting degrees of freedom, and, moreover, is
not diagonalisable in general. 

In order to bring out the salient features of entanglement dynamics,
we consider initial states that are 
direct products of the field and atom modes. Entanglement of these 
two modes sets in during temporal evolution. However, for 
certain ranges of values of the parameters and coupling constants in the 
governing Hamiltonian, we show that the modes disentangle 
at specific instants  during the 
evolution, and the state of the system  
returns close to its initial form (apart from a phase). This 
feature is clearly the 
analogue of wave packet revivals in the dynamics of a 
single mode. Appropriate indicators 
to determine quantitatively the extent and nature of near-revivals and
their fractional counterparts are the  
entanglement entropies, as measured by the 
sub-system von Neumann entropy (SVNE) 
and the sub-system linear entropy (SLE), the sub-system 
considered here being the field mode.  
These entropies show marked dips at near-revivals, and   
are also seen to  reduce 
significantly in certain cases at fractions 
$\frac{1}{2},\, \frac{1}{3}, \,\frac{2}{3}$ and $\frac{1}{4}$ 
of the near-revival time $T_{\rm rev}$, signalling the appearance of 
counterparts  of fractional revivals. 
We have also examined the behaviour of another indicator of
near-revivals, the ``overlap fidelity'' $C(T_{\rm rev})$, 
as a function of a relevant parameter in the
Hamiltonian, namely, the coupling between the sub-systems represented
by the two modes. 
 This  fidelity is defined as the maximum value that the 
overlap $|\inner{\psi(0)}{\psi(t)}|^2$ attains in the vicinity of the 
first near-revival, for a given initial state. 
We study the role of 
the specific (non-entangled) 
initial state considered, and 
the extent of departure of the initial field mode from coherence, 
on the subsequent revival 
properties of the state. In particular, the link 
between the entanglement of states and their  squeezing 
properties has been investigated. 
The initial state of the field mode is taken to be a member of the
family of photon-added coherent states. The advantages of a 
PACS have already been stated. We add that this family of states is
now within the realm of experimental realisation in the 
foreseeable future, 
a single-photon added coherent
state having recently been generated experimentally and characterised 
using quantum tomography\cite{zava}. 
With increasing departure from coherence of the initial field mode, 
the entropy of entanglement at any instant during the temporal evolution 
of the quantum state of the system also increases, and even 
near-revivals do not occur. 

The plan of the rest of this paper is as follows: 
In Section \ref{interaction}, we discuss the relevant features of the 
model Hamiltonian \cite{puri} we use 
to study the dynamics of two-mode entanglement. In Section \ref{entanglement}, we 
examine three different indicators or measures of the
extent to which a state revives at any instant: the first of these    
comprises the instantaneous SVNE and SLE.     
The trends observed here are corroborated by 
the behaviour of the overlap fidelity, which we examine as a function of
the strength of the coupling between the field and the medium.   
Finally, certain operators whose expectation 
values carry signatures of the different fractional revivals 
are also identified. We have 
examined the dynamics of
several initial states, taking the atom to be in the ground 
state while the field is, respectively, in a Fock state, a CS, and a PACS. 
This enables us to analyse systematically  
the effects of different initial field 
modes on the dynamics. The relationship between the 
squeezing property of the state of 
the system and the initial field mode is also brought 
out. 

\section{Single-mode field in a nonlinear medium} 
\label{interaction}
The  interaction of   
a single-mode field of frequency 
$\omega$  with the atoms of the nonlinear medium  through which 
it propagates is modelled by the Hamiltonian\cite{puri} 
\begin{equation}
H =  \omega \,a^\dagger a + \omega_0 \,b^\dagger b +  \gamma\, 
b^{\dagger 2} b^2 +  g \,(a^\dagger b + b^\dagger a).
\label{coupledhamil}
\end{equation}
(We have set $\hbar = 1$.) 
$a$ and $a^\dagger$ are the annihilation and creation operators 
pertaining to the field, while  $b$ and $b^\dagger$ are the
corresponding  atom operators. The medium is modelled 
by an anharmonic oscillator 
with frequency $\omega_0$ and anharmonicity parameter $\gamma$. 
The coupling constant $g$ is a measure 
of the strength of the coupling between the 
field mode and the atom mode. 
It is easily verified that the total number operator 
${\sf N}_{\rm tot} = a^\dagger a + 
b^\dagger b$ commutes with $H$.  We reiterate that the foregoing 
Hamiltonian is not symmetric in the two modes, even when 
$\omega = \omega_0$. 

The Fock basis is given by 
$\{\ket{n'}_a 
\otimes \ket{n}_b\}$, where $n'$ and $n$ are the eigenvalues of  
$a^\dagger a$ and 
$b^\dagger b$, respectively. 
In the absence of the 
coupling constant $g$, $H$ is trivially a direct sum of  
Hamiltonians that are functions of the number operators for the two  
modes. In the absence of 
the anharmonicity parameter $\gamma$, the coupled Hamiltonian is 
essentially linear in each of the sub-system variables, and can be
diagonalised in terms of linear combinations of the original ladder 
operators. In physical terms, this leads to fairly simple dynamics, 
essentially entailing a simple periodic exchange of energy 
between the two sub-systems or modes. When both $g$ and $\gamma$ are
non-zero, the system displays a wide variety of dynamical behaviour, 
depending on the value of the ratio $\gamma/g$. 
Additional insight into the
nature of the model Hamiltonian 
is gained by re-expressing it in terms of angular
momentum operators defined in the usual manner, according to 
$J_{+} = a^{\dagger} b\,,\,\,J_{-} = ab^{\dagger}\,,\,\,
J_z = \frac{1}{2}(a^{\dagger}a - b^{\dagger}b)$. 
We then have  
\begin{equation}
H = \hbar\left[
\textstyle{\frac{1}{2}}(\omega + \omega_0 -\gamma){\sf N}_{\rm tot} 
+ (\omega -\omega_0 + \gamma) J_z 
+ \frac{1}{4}\gamma ({\sf N}_{\rm tot} - 2J_z)^2 
+ 2g J_x\right],
\label{coupledhamil3}
\end{equation}
where 
${\sf N}_{\rm tot}
\left({\sf N}_{\rm tot} +2\right) =4 J^2$.  
As $[{\sf N}_{\rm tot}\,,\,H] = 0$, 
we may write the basis states as 
$\ket{N-n}_a \otimes \ket{n}_b$, 
using $N$ to label the eigenvalues  
of ${\sf N}_{\rm tot}\,$. For notational simplicity, 
let us write   
\begin{equation} 
\ket{N-n}_a \otimes \ket{n}_b \equiv  \ket{N-n\,;\, n}.
\label{prodstate1}
\end{equation} 
It is evident that
$\bra{N-n\,;\, n} H \ket{N'-n'\,;\,n'} = 0$, if $N \neq N'$.  
Hence, for each given value of $N$, 
the Hamiltonian $H$ can be diagonalised 
in the space of the states $\ket{N-n\,;\, n}$, where $n 
= 0,\,1,\,\ldots ,N$. Let the eigenvalues and eigenstates of $H$ be 
$\lambda_{Ns}$ and $\ket{\psi_{Ns}}$, respectively, 
where $s = 0,\,1,\,\ldots ,N$  for a given $N$,
and $N = 0,\,1,\,\ldots \,{\it ad \ inf.} $ 
It is convenient to expand $\ket{\psi_{Ns}}$ 
in the basis $\{\ket{N-n\,;\, n}\}$ as
\begin{equation}
\ket{\psi_{Ns}} = \sum_{n = 0}^{N}\, d_n^{Ns}\,\ket{N-n\,;\, n},
\quad
d_n^{Ns} = \langle{N-n\,;\, n}\,|\,\psi_{Ns}\rangle.
\label{stateexpansion}
\end{equation}
In this basis, (each block of) $H$ 
 can be represented as a real, symmetric, 
tridiagonal matrix. Its eigenvalues 
$\lambda_{Ns}$ and eigenstates $\ket{\psi_{Ns}}$  
can be found numerically using appropriate matrix algebra routines 
\cite{pres,gnu}. An initial state $\ket{\psi(0)}$ of the system 
evolves in time to the state
\begin{equation}
\ket{\psi(t)}=U(t) \ket{\psi(0)}=  \sum_{N=0}^{\infty}\sum_{s=0}^{N}
\,\exp\,(-i\lambda_{Ns}t)\inner{\psi_{Ns}}{\psi(0)}\ket{\psi_{Ns}}
\label{psit}
\end{equation}
at time $t$. 
For our purposes, it is necessary to compute the time-dependent
density operator $\rho(t)$ for  
different choices of the initial state $\ket{\psi(0)}$, as well as 
the reduced density matrices corresponding to the field and atom
sub-systems. 
The main steps in the procedure are outlined in the Appendix.  

As already mentioned, if either $g$ or $\gamma$ is equal to zero, 
the Hamiltonian in Eq. (\ref{coupledhamil}) is exactly 
solvable, and 
there is periodic exchange of energy between the field and atom 
oscillators. 
For non-zero values of the ratio $\gamma/g$ of 
the respective strengths of the nonlinearity and the 
field-atom interaction, collapses and near-revivals could occur 
over certain intervals of time, in between these 
periodic exchanges of energy.  
This phenomenon translates into the behaviour of expectation 
values of certain observables as well. For instance, during  a 
collapse of the energy exchange over an interval of time, the mean 
photon number $\aver{a^\dagger a}$ remains essentially 
constant. A revival of the energy exchange  
is signalled by rapid oscillations of the mean photon number 
about this value, over the relevant time interval. 

Further, when the atomic oscillator is initially in its 
ground state, while the 
field starts  either in a Fock state or in a coherent state, one 
finds the following results\cite{puri}: 
(a)\, For weak nonlinearity ($\gamma/g \ll 1$),  collapses 
and revivals of the mean 
photon number occur almost periodically in time, for both  
kinds of initial field states. The near-revival time is 
approximately given by 
$2\pi/\gamma$ in the former case, and $4\pi/\gamma$ in the 
latter \cite{sesh}. (When $\gamma$ is {\it exactly} equal to zero,
there is no nonlinearity in $H$, and the system is merely
periodic. There is no question of revivals in this case.)  
(b) \,For $\gamma/g \sim 1$, such collapses and revivals 
occur more irregularly if the field is initially in a coherent state,  
compared to the case when it is initially in a Fock state. As the 
 nonlinearity is increased ($\gamma/g \gg 1$),  
collapses and revivals gradually become less discernible.  
Bearing these results in mind, in the next section we examine the 
manner in which the
above-mentioned collapse 
and revival phenomena are mirrored in the 
entropy of entanglement of the system. We  
identify suitable observables which carry signatures of  collapses 
and revivals, and discuss 
the influence of the departure from coherence of 
the initial state of the field 
on the extent to which it revives subsequently. 

\section{Entanglement properties}
\label{entanglement}

We now examine the detailed dynamics of three different initial states 
which are direct products of the field and atom states, evolving under the 
Hamiltonian in Eq. (\ref{coupledhamil}). 
As stated earlier, the initial state of 
the atom  is taken to be the oscillator 
ground state $\ket{0}_b$, while that 
of the field is, respectively, (a) 
a Fock state $\ket{n'}_a\,,\,n' = 0,\,1,\,\ldots \,$;
\, (b) a CS 
$\ket{\alpha}_a\,$; 
and \,(c) an $m$-photon-added CS 
$\ket{\alpha,m}_a\,,\,\,m = 1,\,2,\,\ldots $\,.  
(Recall that the suffixes $a$ and 
$b$ correspond to the electromagnetic field and  
the atoms of the medium, respectively.) The CS and PACS referred to 
have the standard expansions in the Fock basis, namely, 
\begin{equation}
\ket{\alpha} = e^{-|\alpha|^2/2}\sum_{l=0}^{\infty}\frac{\alpha^l}
{\sqrt{l!}}\ket{l}
\label{csdefn}
\end{equation}
and 
\begin{equation}
\added{\alpha}{m}
=\frac{(a^\dagger)^m\ket{\alpha}}{\sqrt{\expect{\alpha}{a^m 
a^{\dagger m}}{\alpha}}}
=\frac{(a^\dagger)^m\ket{\alpha}}{\sqrt{
m!\,L_{m}(-\nu)}},
\label{pacsdefn} 
\end{equation}
where $\alpha \in \mathbb{C}\,,\,\,\nu = |\alpha|^2$, and 
$L_m$ is the Laguerre polynomial of order $m$. 

As we are dealing with a pure bipartite system,  
it is natural to consider the 
time-dependences of $S_k$\,, the sub-system 
von Neumann entropy (SVNE), 
and $\delta_k$\,, the sub-system linear entropy 
(SLE), where the suffix $k$ stands for either $a$ or $b$, 
depending on the sub-system considered. 
These quantities are defined as
\begin{equation}
S_k(t) = -{\rm Tr}_k\,\big\{\rho_k(t) \,\ln \,\rho_k(t)\big\}
\label{svne1}
\end{equation}
and
\begin{equation}
\delta_k(t) = 1 - {\rm Tr}_{k}\,\big\{\rho_k^2(t)\big\},
\label{lne1}
\end{equation}
where $\rho_k(t)$ is the time-dependent reduced 
density operator for 
the sub-system concerned. In terms of the set of 
eigenvalues of $\rho_k(t)$, we have
\begin{equation}
S_k(t) = - \sum_{i} \lambda^{(i)}_{k}(t) \,\ln \,
\lambda^{(i)}_{k}(t),\,\,
\delta_k(t) = 1 - \sum_{i} \big\{\lambda^{(i)}_{k}(t)\big\}^2, 
\label{svnelne}
\end{equation}
where the summation runs over all 
the eigenvalues $\lambda^{(i)}_{k}(t)$.  

\begin{figure}[htpb]
\begin{center}
\includegraphics[width=5.2in]{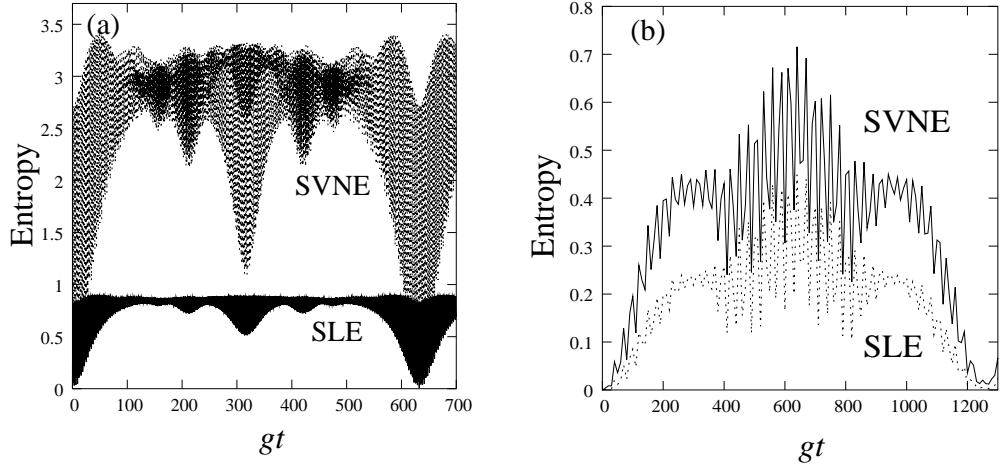}
\caption{SVNE and SLE {\it vs.} $gt$ with $\gamma/g = 10^{-2}$  
for  
(a) an initial Fock state $\ket{10\,;\, 0}$ and 
(b) an initial coherent state $\ket{\alpha\,;\, 0}$ with $\nu = 1$.}
\label{entropy10cross0andalphacross0qbyg.01nu1}
\end{center}
\end{figure}

\begin{figure}[htpb]
\begin{center}
\includegraphics[width=5.2in]
{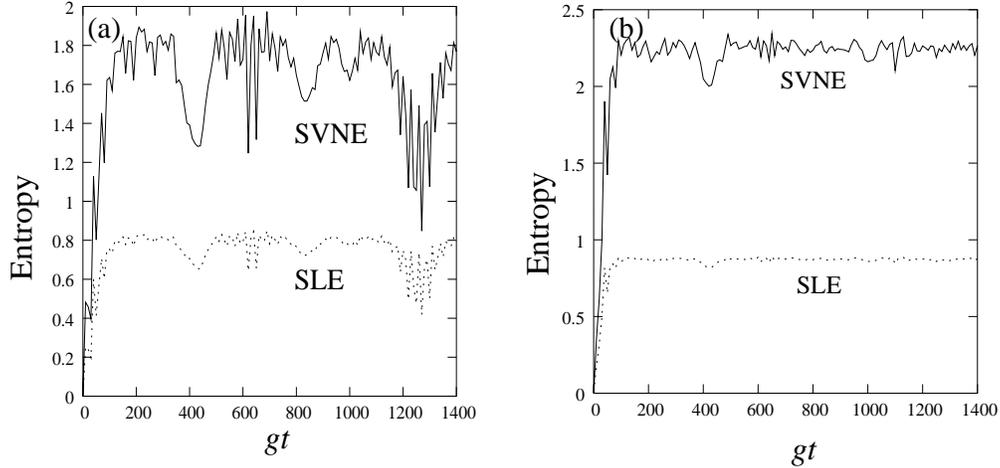}
\caption{SVNE and SLE {\it vs.} $gt$ for an initial 
state $\ket{(\alpha,5)\,;\, 0}$ for (a) $\nu = 1$ and (b) $\nu = 5$\,
($\gamma/g = 10^{-2}$). }
\label{entropyalpha5qbyg.01nu1andnu5}
\end{center}
\end{figure}

Our objective is to investigate the detailed dynamics 
of entangled states \cite{vidal} exhibiting  
revival phenomena. We therefore restrict ourselves here 
to  the case of  weak nonlinearity, i.~e., 
$\gamma/g <<1$, as this is the situation in which 
these phenomena occur most unambiguously. For 
illustrative purposes, we set the values of the
parameters at $\omega = \omega_0 = 1$, and 
$\gamma = 1,\,\,g = 100$, so that   
$\gamma/g = 10^{-2}$. As stated in the preceding section, 
some of the steps involved in the calculation of the foregoing
sub-system entropies are given in the Appendix. 
For pure states in a bipartite system, of course, $S_a = S_b$ and 
$\delta_a = \delta_b$ at any instant of time. Figures 
\ref{entropy10cross0andalphacross0qbyg.01nu1}(a) and (b) 
depict plots of SVNE and SLE versus $gt$ for  
respective initial states  $\ket{10}_a \otimes \ket{0}_b$ 
(or $\ket{10\,;\,0}$, in our notation) 
and  $\ket{\alpha}_a \otimes \ket{0}_b \equiv 
\ket{\alpha\,;\, 
0}$ with the parameter value $\nu = 1$.
The band-like appearance of the plots in 
Fig. \ref{entropy10cross0andalphacross0qbyg.01nu1}(a)  
arises from the extremely rapid oscillations of the ordinates. 
The corresponding plots for an initial state  
$\ket{\alpha,5}_a \otimes  \ket{0}_b \equiv \ket{(\alpha,5)\,;\, 0}$
in the cases $\nu = 1$ and $\nu = 5$ are shown   
in Figs. \ref{entropyalpha5qbyg.01nu1andnu5}(a) and (b) respectively.  
In all the cases above, the SVNE (the upper plot in each figure) 
is larger than the SLE 
(the lower plot in each figure) at any instant of time. 
It is evident that both SVNE and SLE display roughly similar 
oscillatory behaviour in time. 
However, certain striking differences arise in 
the time evolution of the SVNE and 
SLE, depending on the actual initial state considered.
If the field is initially in a Fock state or a CS, 
the entropies return to values close to zero 
at regular intervals of time  
(see Fig. \ref{entropy10cross0andalphacross0qbyg.01nu1}),
signalling a near-revival of the initial state. (Note that 
the revival times are indeed approximately equal to 
$2\pi$ and $4\pi$, respectively, recalling that we have set  
$\gamma = 1$ and $g = 100$.)   
In contrast, if the initial state of the field is a PACS, 
the extent of revival is considerably reduced 
(see Figs. \ref{entropyalpha5qbyg.01nu1andnu5}(a) and (b)).
Further,with an increase in the value of 
$\nu$, the oscillations in the SVNE and SLE die down. This effect is 
enhanced for larger values of $m$, as  seen in  
the rapid increase and saturation of both the SVNE and SLE for an initial 
state $\ket{(\alpha,5)\,;\,0}$ for $\nu = 5$, in contrast 
to the corresponding plots for $\nu = 1$. 

It is also clear that the SVNE and SLE display 
marked oscillatory behaviour near 
$\frac{1}{2}T_{\rm rev}, \frac{1}{3}T_{\rm rev}$ and 
$\frac{1}{4}T_{\rm rev}$. 
This behaviour may be regarded as the counterpart, in our coupled system,
of the fractional revivals seen in the case of a 
single-mode nonlinear Hamiltonian \cite{sudh1,sudh2}. 
Again, these oscillations 
die down in amplitude with increasing 
$m$ when the initial state of the field is a PACS, and are most 
pronounced  if the field is initially in a Fock state. 

\begin{figure}[htpb]
\begin{center}
\includegraphics[width=5.2in]
{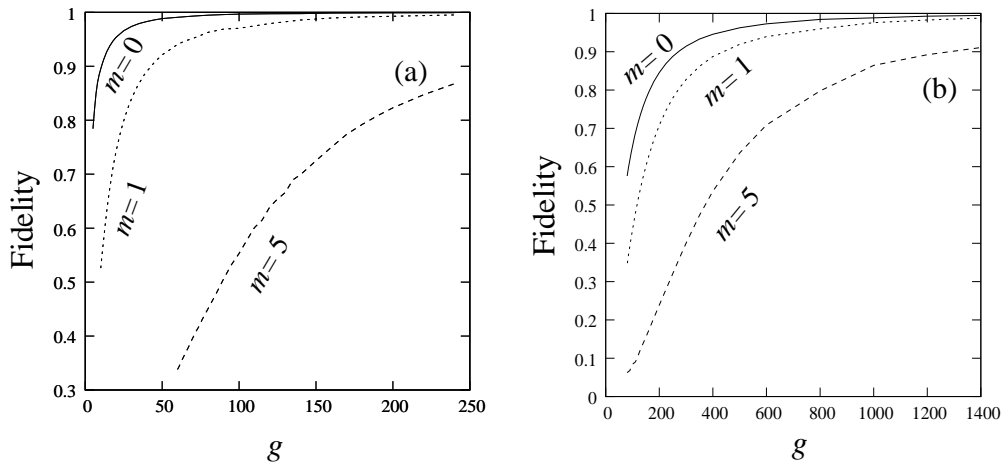}
\caption{Overlap fidelity {\it vs.}  $g$ for an initial 
state $\ket{(\alpha,m)\,;\, 0}$, with 
$\gamma = 1$ and (a) $\nu = 1$ and (b) $\nu = 5$.
}
\label{fidelity}
\end{center}
\end{figure}

As mentioned in Section 1, another 
indicator that characterises the degree of revival of an initial state
is the overlap fidelity, defined as 
\begin{equation}
C(T_{\rm rev}) = \max_{t \,\sim \,T_{\rm rev}} \,
|\inner{\psi(0)}{\psi(t)}|^2,
\label{overlapfid}
\end{equation} 
which is the maximum value attained 
by the overlap function in the vicinity
of the first near-revival time corresponding to a given initial state
$\ket{\psi (0)}$. In Figs. \ref{fidelity}(a) and (b), we have plotted this 
quantity as a function of $g$, the strength of the 
coupling between the field and atom modes, again 
for non-entangled initial states in which
the field is in a CS or a PACS. We see quantitatively how, 
with an increase in the
departure from coherence of the initial field state, near-revivals
occur only for ever increasing values of the coupling strength $g$
relative to the coefficient $\gamma$ of the nonlinearity in the
Hamiltonian.   

Next, we consider whether  
signatures of the features seen above appear 
in the time evolution  
of  expectation values of observables. For this 
purpose, we define the  quadratures  
\begin{equation}
\xi = (x_a + x_b)/2, \quad \eta = (p_a + p_b)/2, 
\label{xiandeta}
\end{equation}
where 
\begin{equation}
x_a = (a + a^\dagger)/\sqrt{2}, \quad x_b = (b + b^\dagger)/\sqrt{2}
\label{xaxb}
\end{equation}
and 
\begin{equation}
p_a = (a - a^\dagger)/(i\sqrt{2}), \quad 
p_b = (b - b^\dagger)/(i \sqrt{2}). 
\label{papb}
\end{equation} 
If the field is initially in a Fock state, both $\aver{\xi}$ and 
$\aver{\eta}$ vanish identically at all times.  
\begin{figure}[htpb]
\begin{center}
\includegraphics[width=5.2in]
{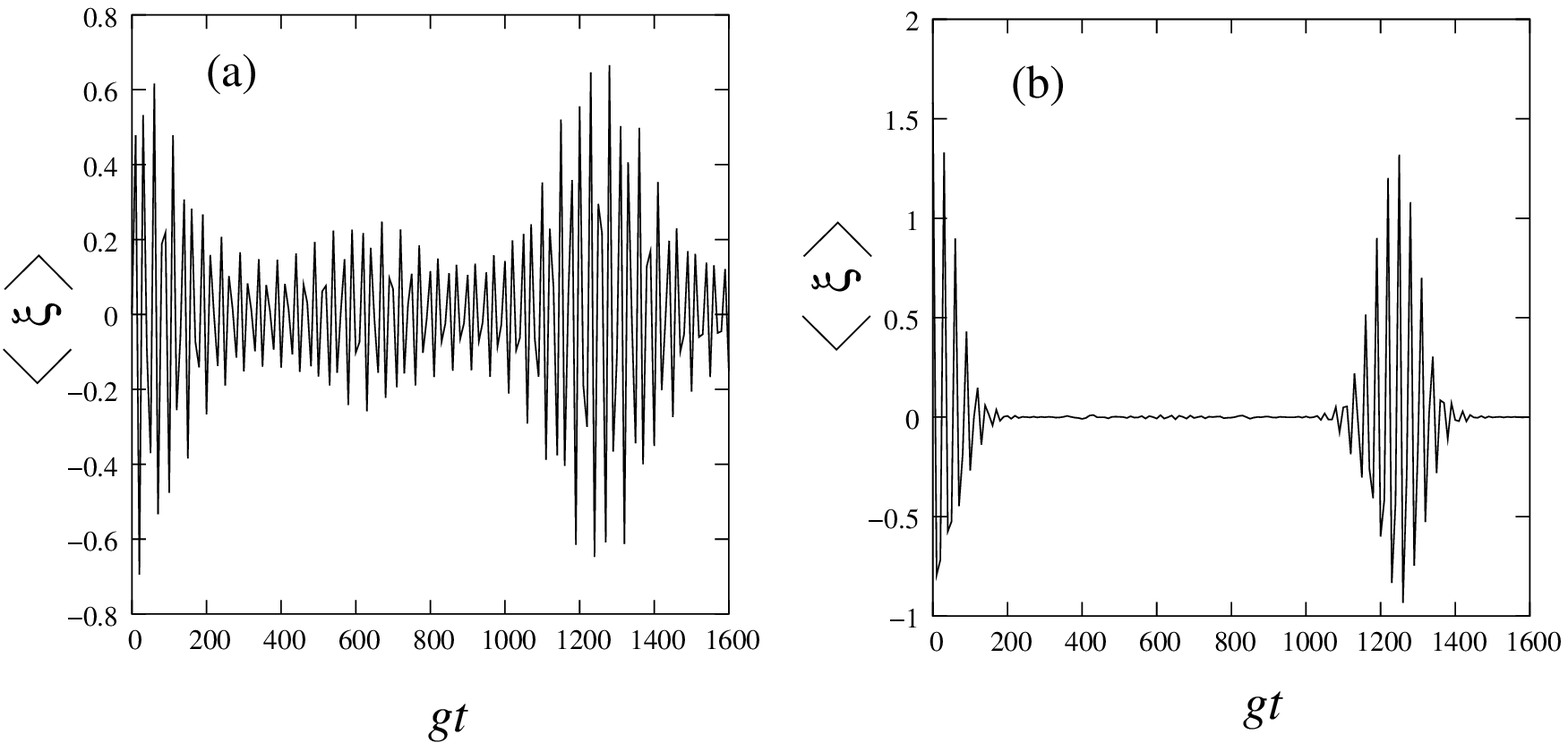}
\caption{$\aver{\xi}$ {\it vs.} $gt$ for an  
initial state $\ket{\alpha; 0}$ 
with and $\nu = 1$ and $\nu = 5$, respectively\,
($\gamma/g 
= 10^{-2}$). }
\label{X1alphacross0qbyg.01nu1andnu5}
\end{center}
\end{figure}
Setting $\omega =\omega_0 = 1,\, \gamma =1$ and $g = 100$ 
as before (for ready comparison with the time evolution of the SVNE 
and SLE discussed above),  
we have plotted $\aver{\xi}$ versus $gt$ for an initial state 
$\ket{\alpha\,;\,0}$ with $\nu = 1$ and $5$, respectively, in Figs.
\ref{X1alphacross0qbyg.01nu1andnu5}(a) and (b).
Figure \ref{X1alphacross0qbyg.01nu1andnu5}(a)  
shows that $\aver{\xi}$  
displays rapid pulsed oscillations near $t = 4\pi$, similar to
its behaviour near $t = 0$. This manifestation of revivals 
is consistent with the behaviour of 
the SVNE and SLE in this case, {\it cf.} Fig. 
\ref{entropy10cross0andalphacross0qbyg.01nu1}(b). The collapses 
are not sharp, in the 
sense that $\aver{\xi}$ is not constant over the time interval between 
successive revivals---oscillatory bursts 
occur in between, with a slight enhancement of these
oscillations around the fractional revival at 
$\frac{1}{2}T_{\rm rev}$.  
In contrast, the collapses in between revivals 
are much more
complete for larger values of $\nu$, as seen in Fig. 
\ref{X1alphacross0qbyg.01nu1andnu5}(b), consistent with 
the corresponding behaviour of the SVNE and SLE 
in this case.  
\begin{figure}[htpb]
\begin{center} 
\includegraphics[width=2.5in]
{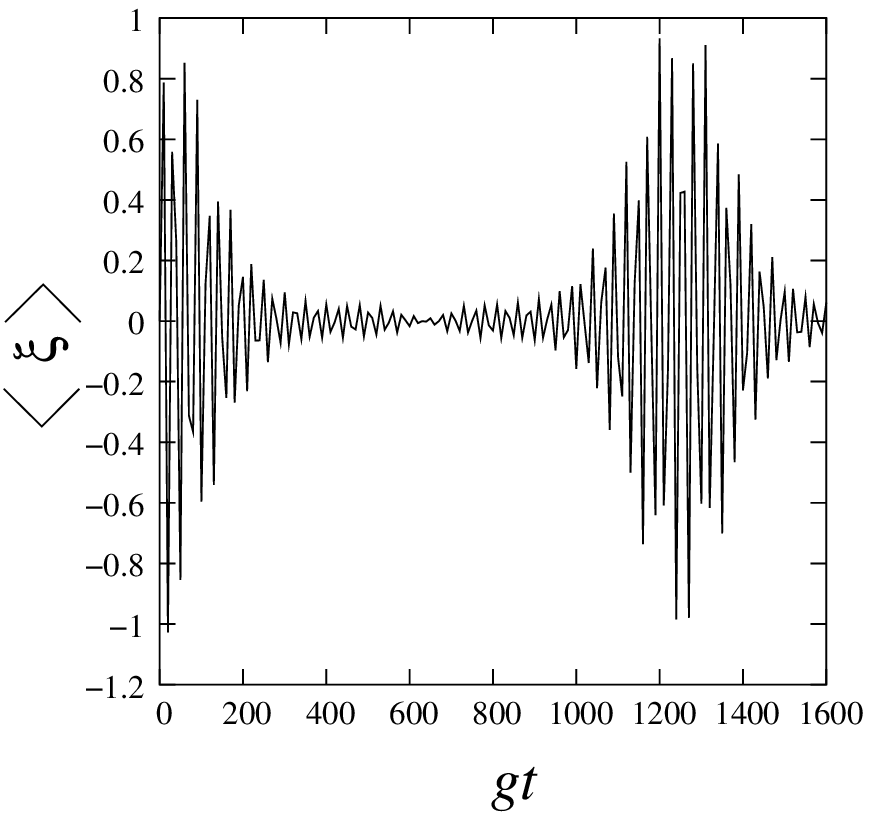}
\caption{$\aver{\xi}$ {\it vs.} $gt$ for 
an initial state  
$\ket{(\alpha,1)\,;\,0}$ with 
$\nu = 1$\, ($\gamma/g = 10^{-2}$).}
\label{X1alpha1cross0qbyg.01nu1}
\end{center} 
\end{figure}
An interesting feature is that 
these collapses become much sharper
for even a marginal departure from 
coherence of 
the initial state of the field (i.~e., even for as low a value as 
$m = 1$).  $\aver{\xi}$ remains  
virtually constant 
over the duration of the collapse, 
and then bursts into rapid oscillations 
close to revivals, as seen in 
Fig. \ref{X1alpha1cross0qbyg.01nu1} 
which corresponds to 
$m = 1$ and $\nu = 1$. As in the case of 
single-mode dynamics, 
the amplitude 
of the oscillations in the neighbourhood of $T_{\rm rev}$  
decreases 
significantly with an increase in $m$.  
Thus, for small values of 
$\nu$, it is easy to distinguish between 
an initial CS and an initial PACS. The 
expectation value of  $\eta$ also displays these 
signatures. We may add that, while the 
sub-system variables $x_a$, $x_b$, $p_a$ and 
$p_b$ do exhibit near-revivals in their 
expectation values, their higher moments 
do not capture the occurrence of 
fractional revivals. However, 
the higher moments of the combinations 
$\xi$ and $\eta$ do 
carry distinguishing signatures to 
selectively pin-point the analogues of the 
different fractional revivals that 
occur in the single-mode case. Hence $\xi$ and $\eta$ are 
the appropriate dynamical variables in the interacting  
system under consideration. 

\begin{figure}[htpb]
\begin{center}
\includegraphics[width=2.5in]
{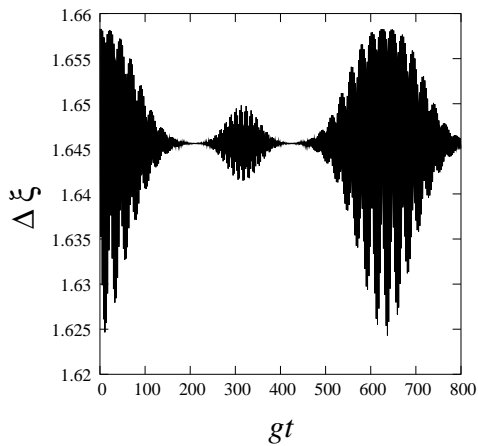}
\caption{$\Delta \xi$ {\it vs.} $gt$ for an initial state 
$\ket{10\,;\,0}$\,  
($\gamma/g = 10^{-2}$).}
\label{deltaX110cross0qbyg.01}
\end{center}
\end{figure}

\begin{figure}[htpb]
\begin{center}
\includegraphics[width=2.5in]
{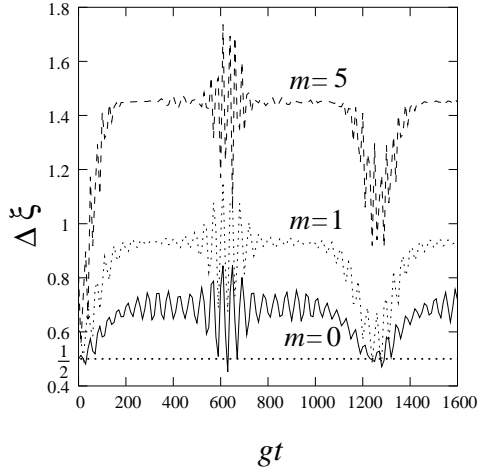}
\caption{$\Delta \xi$ {\it vs.} $gt$ for an initial 
state $\ket{(\alpha,m)\,;\, 0}$ with
$\nu = 1$ and $m = 0, 1$ and $5$, respectively\,
($\gamma/g = 10^{-2}$).}
\label{deltaX1alphacomma01and5cross0qbyg.01nu1}
\end{center}
\end{figure}

The standard deviation $\Delta{\xi}$ of $\xi$ 
reflects the occurrence of the dips 
in the plots of the 
SVNE and SLE at $\frac{1}{2}T_{\rm rev}$. 
The plot of $\Delta{\xi}$ versus $gt$ for an initial state 
$\ket{10\,;\, 0}$ shows a burst of rapid oscillations at 
$t \simeq \pi$  (Fig. \ref{deltaX110cross0qbyg.01}). 
This feature holds for an initial CS or  PACS as well,  
as is evident (Fig. 
\ref{deltaX1alphacomma01and5cross0qbyg.01nu1})  
from the  sudden burst of 
oscillations in $\Delta \xi$ 
around $t \simeq 2\pi$ 
for initial states  
$\ket{\alpha\,;\, 0}$,
$\ket{(\alpha,1)\,;\, 0}$ and $\ket{(\alpha,5)\,;\, 0}$
(recall that $T_{\rm rev} \simeq 4\pi$ in this case). 

We note that {\it squeezing} occurs in the neighbourhood of  
$\frac{1}{2}T_{\rm rev}$
when the initial state 
of the field is a coherent 
state: $\Delta \xi$ drops below the 
value $\frac{1}{2}$ (the horizontal
dotted line in 
Fig. \ref{deltaX1alphacomma01and5cross0qbyg.01nu1}) 
in the case when the initial state is 
$\ket{\alpha \,;\, 0}$,  
in contrast to what happens for an initial 
PACS $\ket{(\alpha,m)\,;\,0}$. 
While this is similar to  
the squeezing property seen in the  
case of single-mode dynamics, such parallels 
do not hold in the case of higher-order 
squeezing. The relevant quadrature variables 
in this case are obvious generalisations of those 
considered in the case of single-mode dynamics \cite{du}, 
and are given by \cite{sudh4} 
\begin{equation}
Z_1=\frac{(a^q+a^{\dagger q}+b^q+b^{\dagger q})}{2\sqrt{2}},\quad
Z_2=\frac{(a^q-a^{\dagger q}+b^q-b^{\dagger q})}{2i\sqrt{2}}.
\label{Z1Z2coupled}
\end{equation}
In contrast to the single-mode example, 
even for weak nonlinearity ($\gamma/g = 
10^{-2}$) and $\nu = 1$, amplitude-squared 
squeezing ($q = 2$) is absent at $t 
= \frac{1}{2}T_{\rm rev}$ whether the 
field is in a Fock state, or a CS, or a PACS. 

Finally, turning to the higher moments of 
$\xi$ and $\eta$, we note that all odd 
moments of $\xi$ vanish identically for all $t$ 
if the initial state is a direct product of Fock states.
For small values of $\nu$,  
the higher moments  of $\xi$ 
show distinct signatures 
at fractional 
revivals only if $m$ is sufficiently 
large. However, for larger values of $\nu$, 
such signatures appear even in the 
case of an initial CS ($m = 0$). 
In contrast to this, the temporal evolution of the 
variances and higher 
moments of the subsystem variables $x_a$, $x_b$, $p_a$ and $p_b$ do not 
display these signatures. In this sense,  
the expectation values of sub-system 
quadrature variables would seem to be
inappropriate choices for 
investigating collapse and revival 
phenomena in the presence of entanglement, although, 
as we have shown, the sub-system entropies are 
eminently suitable indicators in this regard.  
 
The model Hamiltonian we have used to study two-mode dynamics 
leads to several other interesting
features that are manifested in the expectation values of sub-system
variables, as the parameters in $H$ are varied. In particular, the
ergodicity properties of the system, in a ``phase space'' spanned by
such expectation values, exhibit a range of behaviour from
quasi-periodicity to exponential instability---the latter, 
notwithstanding the fact that the classical counterpart of $H$ is an
integrable two-freedom Hamiltonian. 
These results will be reported elsewhere.

\acknowledgments
This work was supported in part by the
Department of Science and Technology, India, under Project No.
SP/S2/K-14/2000.

\appendix
\section*{Appendix: Calculation of the density matrix}
\setcounter{section}{1}

We outline here the procedure used for calculating the density 
matrix elements required for the determination 
of the entropies and
related quantities in the main text.  

From Eq. (\ref{psit}) for the state vector of the system at time $t$,
it follows that the time-dependent density matrix $\rho(t)$ is 
\begin{eqnarray}
\rho(t)=\sum_{N=0}^{\infty}\sum_{s=0}^{N}
\sum_{N'=0}^{\infty}\sum_{s'=0}^{N'}
&\exp\,[-i(\lambda_{Ns}-\lambda_{N's'})\,t\,]\nonumber \\
&\times \inner{\psi_{Ns}}{\psi(0)}\inner{\psi(0)}{\psi_{N's'}}
\ket{\psi_{Ns}}\bra{\psi_{N's'}}.
\label{rhotgeneral}
\end{eqnarray}
For instance, if the atomic oscillator is initially 
in the ground state 
$\ket{0}_b$ 
and the field is in the Fock 
state $\ket{N}_a$, the density matrix reduces to 
\begin{equation}
\rho(t) = \sum_{s = 0}^{N}\,\sum_{s' = 
0}^{N}\,\exp\,[-i(\lambda_{Ns} - \lambda_{Ns'})t]
\,\,d_0^{Ns}\,\,d_0^{Ns'}\,
\ket{\psi_{Ns}}\bra{\psi_{Ns'}},
\label{rhotN}
\end{equation} 
in   terms of the expansion coefficients $d_n^{Ns}$ defined in
Eq. (\ref{stateexpansion}).  
(This expression is explicitly $N$-dependent, 
as expected.) Hence
\begin{eqnarray}
\expect{\psi_{Ml}}{\rho(t)}{\psi_{M'l'}}=
\,\exp\,[-i(\lambda_{Ml}-\lambda_{M'l'})\,t\,]\,\,
d_0^{Ml}\,\,d_0^{M'l'}\,\delta_{MN}\delta_{NM'}.
\end{eqnarray} 
It is evident that $\rho(t)$ is effectively an 
$(N+1)$-dimensional diagonal matrix in this case. \\

For an initial state $\ket{(\alpha,m); 0}$ we find, using
Eq. (\ref{rhotgeneral}),
\begin{eqnarray}
\rho(t)&=&\frac{e^{-\nu}}{m!L_m(-\nu)}\sum_{N=m}^{\infty}\sum_{s=0}^{N}
\sum_{N'=m}^{\infty}\sum_{s'=0}^{N'}
\frac{\sqrt{N!\,N'!}\,\,(\alpha)^{N-m}\,\,
(\alpha^*)^{N'-m}}{(N-m)!\,(N'-m)!}\nonumber\\
&\times&\,\exp\,[-i(\lambda_{Ns}-\lambda_{N's'})\,t\,]\,
d_0^{Ns}\,\,d_0^{N's'}\,\ket{\psi_{Ns}}\bra{\psi_{N's'}},
\end{eqnarray}
where we have used the expansion of the PACS  
$\ket{\alpha, m}$ in the Fock basis.
The corresponding matrix elements of the density matrix are given by
\begin{eqnarray}
\expect{\psi_{Ml}}{\rho(t)}{\psi_{M'l'}}&=&\frac{e^{-\nu}}
{m!L_m(-\nu)}\frac{\sqrt{M!\,M'!}\,\,(\alpha)^{M-m}\,\,(\alpha^*)^{M'-m}}
{(M-m)!\,(M'-m)!}\nonumber\\ 
&\times&
\,\exp\,[-i(\lambda_{Ml}-\lambda_{M'l'})\,t\,]\,d_0^{Ml}\,\,d_0^{M'l'}.
\end{eqnarray}
Here, and in the rest of this Appendix, it is understood that 
contributions 
from terms of the form $1/(-n)!$, where $n$ is a positive integer, vanish.

The expectation values and higher moments of the quadrature 
variables $\xi(t)$ and $\eta(t)$, defined in Eq. (\ref{xiandeta}), 
can now be 
obtained numerically, using the 
above 
expressions for the density matrix and the
matrix elements of the operators 
$a$ and $b$ in the basis $\ket{\psi_{Ns}}$. The latter are given by
\begin{equation}
\expect{\psi_{Ns}}{a}{\psi_{N's'}}=
\sum_{n=0}^{N'} (N'-n)^{1/2}\,\,d_{n}^{Ns}\,d_{n}^{N's'}\,\delta_{N,N'-1}
\label{matrixa}
\end{equation}
and
\begin{equation}
\expect{\psi_{Ns}}{b}{\psi_{N's'}}=
\sum_{n=1}^{N'} n^{1/2}\,\,d_{n-1}^{Ns}\,d_{n}^{N's'}\,\delta_{N,N'-1}\,,
\label{matrixb}
\end{equation}
respectively. As these are purely off-diagonal, and $\rho(t)$ is 
diagonal for an initial state that is a direct product of Fock states, 
it follows that all the odd moments of $\xi$ and $\eta$ vanish identically
for all $t$, as asserted in the text. This is no longer 
true for the other classes of initial states considered.
 
The time-dependent reduced density matrices $\rho_k(t)$ ($k = a,b)$ 
are given by
\begin{eqnarray}
\rho_a(t)&=&{\rm Tr}_b\,[\rho(t)]=
\sum_{n=0}^{\infty}
\,_b\!\expect{n}{\rho(t)}{n}_{b}\,,\nonumber\\
\rho_b(t)&=&{\rm Tr}_a\,[\rho(t)]=
\sum_{n=0}^{\infty}\,_a\!\expect{n}{\rho(t)}
{n}_{a}.
\end{eqnarray}
Corresponding to an initial state $\ket{N\,;\, 0}$, 
these reduced density matrices $\rho_k(t)$ take the form
\begin{eqnarray}
\rho_a(t) &=& \sum_{n=0}^{N}\sum_{s = 0}^{N}\,\sum_{s' = 
0}^{N}\,\exp\,[-i(\lambda_{Ns} - \lambda_{Ns'})\,t\,]\nonumber \\
&&\times \quad d_0^{Ns}\,\,d_0^{Ns'}\,\,d_n^{Ns}\,\,d_n^{Ns'}\,
\ket{(N-n)}_{a} \,_a\!\bra{(N-n)}
\label{}
\end{eqnarray}
and
\begin{eqnarray}
\rho_b(t) &=& \sum_{n=0}^{N}\sum_{s = 0}^{N}\,\sum_{s' = 
0}^{N}\,\exp\,[-i(\lambda_{Ns} - \lambda_{Ns'})\,t\,]\nonumber \\
&&\times \quad  d_0^{Ns}\,\,d_0^{Ns'}\,\,d_{N-n}^{Ns}\,\,d_{N-n}^{Ns'}\,
\ket{(N-n)}_b\,_b\!\bra{(N-n)}.
\label{}
\end{eqnarray}
Hence we have, in the Fock basis, 
\begin{eqnarray}
_a\!\expect{n}{\rho_{a}(t)}{n'}_a &=& \sum_{s = 0}^{N}\,\sum_{s' = 
0}^{N}\,\exp\,[-i(\lambda_{Ns} - \lambda_{Ns'})\,t\,]\,
\,d_0^{Ns}\,\,d_0^{Ns'}\,\,d_{N-n}^{Ns}\,\,d_{N-n'}^{Ns'}\,\delta_{nn'}
\nonumber\\
\label{subrhoatfock}
\end{eqnarray}
and
\begin{eqnarray}
_b\!\expect{n}{\rho_b(t)}{n'}_b &=& \sum_{s = 0}^{N}\,\sum_{s' = 
0}^{N}\,\exp\,[-i(\lambda_{Ns} - \lambda_{Ns'})\,t\,]\,
\,d_0^{Ns}\,\,d_0^{Ns'}\,\,d_{n}^{Ns}\,\,d_{n'}^{Ns'}\,\delta_{nn'}\,
.\nonumber\\
\label{subrhobtfock}
\end{eqnarray}
As before, these are explicitly $N$-dependent finite-dimensional 
matrices. 

For an initial 
state 
$\ket{(\alpha,m)\,;\, 0}$  
the expressions for $\rho_k(t)$ are given by
\begin{eqnarray}
\rho_a(t)&=&\frac{e^{-\nu}}{m!L_m(-\nu)}
\sum_{n=0}^{\infty}\sum_{N=N_{\rm min}}^{\infty}\sum_{s=0}^{N}
\sum_{N'=N_{\rm min}}^{\infty}
\sum_{s'=0}^{N'}
\frac{\sqrt{N!\,N'!}\,\,(\alpha)^{N-m}\,\,(\alpha^*)^{N'-m}}
{(N-m)!\,(N'-m)!}\nonumber\\
&\times&\,\exp\,[-i(\lambda_{Ns}-\lambda_{N's'})\,t\,]\,
d_0^{Ns}\,\,d_0^{N's'}\,\,d_n^{Ns}\,\,d_n^{N's'}\,\,\ket{(N-n)}_a\,
_a\!\bra{(N'-n)}\nonumber\\
\end{eqnarray}
and
\begin{eqnarray}
\rho_b(t)&=&\frac{e^{-\nu}}{m!L_m(-\nu)}\sum_{n=0}^{\infty}
\sum_{N=N_{\rm min}}^{\infty}\sum_{s=0}^{N}\sum_{N'=N_{\rm min}}^{\infty}
\sum_{s'=0}^{N'}
\frac{\sqrt{N!\,N'!}\,\,(\alpha)^{N-m}\,\,(\alpha^*)^{N'-m}}
{(N-m)!\,(N'-m)!}\nonumber\\
&\times&
\exp\,[-i(\lambda_{Ns}-\lambda_{N's'})\,t\,]\,
d_0^{Ns}\,\,d_0^{N's'}\,d_{N-n}^{Ns}\,\,d_{N'-n}^{N's'}\,\,
\ket{(N-n)}_b\,
_b\!\bra{(N'-n)},\nonumber\\
\end{eqnarray}
where $N_{\rm min}=\max\,(n,m)$. These are infinite-dimensional 
matrices. 
The corresponding matrix elements of $\rho_k(t)$  
in the Fock basis are given by 
\begin{eqnarray}
_a\!\expect{l}{\rho_a(t)}{l'}_a&=&\frac{e^{-\nu}}{m!L_m(-\nu)}
\sum_{n_{\rm min}}^{\infty}\sum_{s=0}^{n+l}
\sum_{s'=0}^{n+l'}
\frac{\sqrt{(n+l)!\,(n+l')!}\,\,(\alpha)^{n+l-m}\,\,(\alpha^*)^{n+l'-m}}
{(n+l-m)!
\,(n+l'-m)!}\nonumber\\
&\times&
\,\exp\,[-i(\lambda_{(n+l)s}-\lambda_{(n+l')s'})\,t\,]\,
\,d_0^{(n+l)s}\,\,d_0^{(n+l')s'}\,\,
d_n^{(n+l)s}\,\,d_n^{(n+l')s'}\nonumber\\
\label{subrhoatphoton}
\end{eqnarray}
and
\begin{eqnarray}
_b\!\expect{l}{\rho_b(t)}{l'}_b&=&\frac{e^{-\nu}}{m!L_m(-\nu)}
\sum_{n_{\rm min}}^{\infty}\sum_{s=0}^{n+l}
\sum_{s'=0}^{n+l'}
\frac{\sqrt{(n+l)!\,(n+l')!}\,\,(\alpha)^{n+l-m}\,\,(\alpha^*)^{n+l'-m}}
{(n+l-m)!
\,(n+l'-m)!}\nonumber\\
&\times&
\,\exp\,[-i(\lambda_{Ns}-\lambda_{N's'})\,t\,]\,\,
d_0^{(n+l)s}\,\,d_0^{(n+l')s'}\,\,d_{l}^{(n+l)s}\,\,d_{l'}^{(n+l')s'},\nonumber\\
\label{subrhobtphoton}
\end{eqnarray}
where $n_{\rm min}=\max\,(m-l,m-l')$. The corresponding expressions in
the case of an initial state 
$\ket{\alpha\,;\, 0}$ are obtained from the above by simply setting $m
= 0$.   

All the reduced density matrices (with elements 
given by Eqs. (\ref{subrhoatfock}), 
(\ref{subrhobtfock}), (\ref{subrhoatphoton}) 
and (\ref{subrhobtphoton})) 
are hermitian. They are diagonalised numerically, and their eigenvalues 
are used to compute the entropies $S_k(t)$ and $\delta_k(t)$ given 
by Eqs. (\ref{svnelne}). In the case of infinite-dimensional 
matrices, rapid convergence in numerical computation 
is provided by the factorials 
in the denominators of the summands in the expressions derived above for 
the matrix  elements. We use double precision arithmetic with an accuracy 
of $1$ part in $10^{6}$.  As mentioned in the text, we use the equality of
$S_a(t)$ and $S_b(t)$, and the condition $\rho^2(t)=\rho(t)$ 
for the density matrix of the total system, 
as some of the checks on the numerical 
computations, as we are only dealing with pure states.

\end{document}